\newcommand{\onlytsatt}{\textup{\textsc{3sat}}}
\newcommand{\onlysat}{\textup{\textsc{sat}}}
\newcommand{\onlyand}{\textup{\textsc{AND}}}
\newcommand{\onlyor}{\textup{\textsc{OR}}}
\newcommand{\onlynot}{\textup{\textsc{NOT}}}

\documentclass[10pt,conference]{IEEEtran}
\IEEEoverridecommandlockouts
\usepackage{cite}
\usepackage{amsmath,amssymb,amsfonts, amsthm}

\usepackage{subcaption}
\usepackage{graphicx}
\usepackage{textcomp}
\usepackage{todonotes}
\usepackage{xcolor}
\usepackage{algpseudocode}

\def\BibTeX{{\rm B\kern-.05em{\sc i\kern-.025em b}\kern-.08em
    T\kern-.1667em\lower.7ex\hbox{E}\kern-.125emX}}

\DeclareMathOperator*{\argmin}{arg\,min}

\theoremstyle{break}
\newtheorem{dfn}{Definition}
\theoremstyle{definition}
\newtheorem{exmp}{Example}

\begin{document}
  
\title{Pattern QUBOs: Algorithmic construction of \onlytsatt{}-to-QUBO transformations}

\author{\IEEEauthorblockN{ Sebastian Zielinski}
\IEEEauthorblockA{\textit{Institute for Infomatics} \\
\textit{LMU Munich}\\
Munich, Germany \\
sebastian.zielinski@ifi.lmu.de}
\and
\IEEEauthorblockN{Jonas Nüßlein}
\IEEEauthorblockA{\textit{Institute for Infomatics} \\
\textit{LMU Munich}\\
Munich, Germany \\
jonas.nuesslein@ifi.lmu.de}
\and
\IEEEauthorblockN{Jonas Stein}
\IEEEauthorblockA{\textit{Institute for Infomatics} \\
\textit{LMU Munich}\\
Munich, Germany \\
jonas.stein@ifi.lmu.de}
\and
\IEEEauthorblockN{Thomas Gabor}
\IEEEauthorblockA{\textit{Institute for Infomatics} \\
\textit{LMU Munich}\\
Munich, Germany \\
thomas.gabor@ifi.lmu.de}
\and
\IEEEauthorblockN{Claudia Linnhoff-Popien}
\IEEEauthorblockA{\textit{Institute for Infomatics} \\
\textit{LMU Munich}\\
Munich, Germany \\
linnhoff@ifi.lmu.de}
\and
\IEEEauthorblockN{Sebastian Feld}
\IEEEauthorblockA{\textit{Quantum \& Computer Engineering} \\
\textit{Delft University of Technology}\\
Delft, The Netherlands \\
s.feld@tudelft.nl}
}

\maketitle
\begin{abstract}
\onlytsatt{} instances need to be transformed into instances of Quadratic Unconstrained Binary Optimization (QUBO) to be solved on a quantum annealer. Although it has been shown that the choice of the \onlytsatt{}-to-QUBO transformation can impact the solution quality of quantum annealing significantly, currently only a few \onlytsatt{}-to-QUBO transformations are known. Additionally, all of the known \onlytsatt{}-to-QUBO transformations were created manually (and not procedurally) by an expert using reasoning, which is a rather slow and limiting process. In this paper, we will introduce the name \emph{Pattern QUBO} for a concept that has been used implicitly in the construction of \onlytsatt{}-to-QUBO transformations before. We will provide an in-depth explanation for the idea behind Pattern QUBOs and show its importance by proposing an algorithmic method that uses Pattern QUBOs to create new \onlytsatt{}-to-QUBO transformations automatically.
As an additional application of Pattern QUBOs and our proposed algorithmic method, we introduce \emph{approximate \onlytsatt{}-to-QUBO transformations}. These transformations sacrifice optimality but use significantly fewer variables (and thus physical qubits on quantum hardware) than non-approximate \onlytsatt{}-to-QUBO transformations. We will show that approximate \onlytsatt{}-to-QUBO transformations can nevertheless be very effective in some cases.
\end{abstract}



\section{Introduction}\noindent Satisfiability problems occupy a central place in computer science. They have been amongst the first problems, for which NP-completeness has been shown \cite{cook1971complexity} and they have since often been used to proof the NP-completeness of other NP problems \cite{arora2009computational}. Informally the satisfiablilty problem (\onlysat) of propositional logic is defined as follows: Given a Boolean formula, is there any assignment of Boolean values to the involved variables such that the formula evaluates to ``true''? \cite{nusslein2023solving}. These problems occur in various different domains like planning \cite{kautz1992planning}, artificial intelligence \cite{arora2009computational}, formal verification \cite{prasad2005survey}, automatic test-pattern generation \cite{marques2008practical}, and more. Thus, in the past decades several different methods for solving satisfiability problems have been developed \cite{schoning1999probabilistic, selman1993local, marques2003grasp, davis1962machine, kurin2019improving}. Despite these efforts, to this day no algorithm is known that can solve any instance of satisfiability problems in worst-case polynomial time.\\
Quantum computing, however, is a computational model that can solve some NP problems exponentially faster \cite{shor1994algorithms} than their classical counterparts, by using phenomena of quantum mechanics for their calculations. There are two main approaches to quantum computing: the quantum gate model and adiabatic quantum computing. One method of solving hard problems with quantum computers is to transform them into an instance of Quadratic Unconstrained Binary Optimization (QUBO) or into an instance of an Ising spin glass problem (Ising), because QUBO and Ising can be used as an input to quantum annealers \cite{mcgeoch2020theory} or the quantum approximate optimization algorithm (QAOA \cite{farhi2014quantum}) on quantum gate computers.\\\\ Due to recent advances in the manufacturing of quantum computers, the research community directs huge effort to employing quantum techniques to (NP-)hard problems in their respective fields. In the recent past, a particular research effort was to transform NP-hard optimization problems into instances of QUBO or Ising (see \cite{lucas2014ising}, e.g., for an overview). It is thus not surprising that the research community also studied satisfiability problems in the context of quantum computing (see \cite{nusslein2022algorithmic, gabor2019assessing, sax2020approximate}, e.g.). To this day several methods of transforming satisfiability problems to instances of QUBO have been published (see \cite{choi2010adiabatic, chancellor2016direct, nusslein2023solving}, e.g.). Probably all of these \onlytsatt{}-to-QUBO transformations have in common that they were carefully crafted  by experts (i.e., manually and not by a procedural method) in this field. In consideration of several studies that show that the choice of a \onlytsatt{}-to-QUBO transformation can significantly impact the solution quality of quantum annealing (\cite{kurin2019improving, zielinski2023influence}), we argue that creating these transformations by hand does not seem to be sufficient. The space of potential \onlytsatt{}-to-QUBO transformations might be large (although we do not know how large exactly) and the number of \onlytsatt{}-to-QUBO transformations in that space that can be discovered by an expert using clever reasoning seems to be limited.\\

\noindent In this paper we propose an algorithmic method that automatically creates new \onlytsatt{}-to-QUBO transformations. Our algorithmic method builds upon a concept that has been used in the construction of known \onlytsatt{}-to-QUBO transformations (the most explicit description of this concept to this day can be found in \cite{nusslein2023solving}). Although this concept seems to be already in use, due to its importance for the construction of algorithms that are able to create new \onlytsatt{}-to-QUBO transformations automatically, we argue that this concept needs to be pointed out even more explicitly. Thus, our second contribution is to explicitly introduce the name \emph{pattern QUBOs} for this concept, provide an in-depth description for it and show its use in the creation of algorithms that automatically create \onlytsatt{}-to-QUBO transformations. As our third and final contribution we will introduce \emph{approximate \onlytsatt{}-to-QUBO transformations} as an additional application of our algorithmic method. These transformations sacrifice optimality (correctness) but need significantly fewer variables (and thus significantly fewer qubits on quantum hardware) than non-approximate \onlytsatt{}-to-QUBO transformations. We show that approximate \onlytsatt{}-to-QUBO transformations can be surprisingly effective. \\

\noindent The remainder of this paper is structured as follows: Section \ref{sec:background} introduces necessary foundations for this paper, namely satisfiability problems, QUBO, and the maximum-weight independent set problem MWIS. In Section \ref{sec:soa_qubos}, we review state-of-the-art \onlytsatt{}-to-QUBO transformations that will later be used in a case study as a comparison. In Section \ref{sec:pattern_qubos}, we formally introduce pattern QUBOs and explain the concept of pattern QUBOs in depth. In Section \ref{sec:algorithm}, we present our algorithmic method to finding new \onlytsatt{}-to-QUBO transformations automatically and compare one of the new \onlytsatt{}-to-QUBO transformations to state-of-the-art \onlytsatt{}-to-QUBO transformations in a small case study on D-Wave's quantum annealer Advantage\_system4.1. In Section \ref{sec:approximate_3sat}, we define \emph{approximate \onlytsatt{}-to-QUBO transformations} and show that they can be surprisingly effective in a case study on D-Wave's quantum annealer Advantage\_system4.1 and on classical hardware using D-Wave's Tabu Sampler. Finally we conclude the paper in Section \ref{sec:conclusion} and state further research topics.


\section{Background}\label{sec:background}


\subsection{Satisfiability Problems}\label{sec:sat_problems}
\noindent Satisfiability problems are concerned with solving special types of Boolean formulae. Thus we now formally introduce Boolean formulae as presented in \cite{arora2009computational}. \\

\begin{dfn}[Boolean formula]
Let $x_1, ..., x_n \in \{0, 1\}$ be Boolean variables. We set $0 := False$ and $1 := True$. A Boolean formula  $\varphi$ over $n$ Boolean variables consists of the variables $x_1, ..., x_n$ and the logical operators \onlyand ($\wedge$), \onlyor ($\vee$), \onlynot ($\neg$).
For a given $z \in \{0, 1\}^n$ the expression $\varphi(z)$ denotes the value of $\varphi$ when each of the Boolean variables $x_i$ is assigned the value $z_i$. If there exists some assignment $z$ such that $\varphi(z) = 1 (True)$ we call $\varphi$ satisfiable. Otherwise we call $\varphi$ unsatisfiable.\cite{arora2009computational}
\end{dfn}

\noindent In this work we are adressing \onlytsatt{} problems, which are concerned with the satisfiability of Boolean formulae of a specific structure. The specific required structure is called \emph{Conjunctive Normal Form (CNF)}:

\begin{dfn}[Conjunctive Normal Form]
\theoremstyle{break}
Let $\varphi$ be a Boolean formula over the Boolean variables $x_1, ..., x_n$. We say that $\varphi$ is in CNF form, if $\varphi$ is of the following form:
$$
\bigwedge_{i}\big( \bigvee_j l_{i{_j}}\big)
$$
The terms $l_{i{_j}}$ are called literals. The value of a literal is either a variable $x_k$ or its negation $\neg x_k$ (for $k \in \{1, ..., n\})$. The terms $\big( \bigvee_j l_{i{_j}}\big)$ are called clauses. A \emph{kCNF} is a formula in CNF form in which all clauses contain at most $k$ literals.
\end{dfn}

\noindent We can now formally define \onlytsatt{} problems:

\begin{dfn}[3SAT]
A \onlytsatt{} instance is a Boolean formula in 3CNF form, which comprises $n$ Boolean variables and $m$ clauses. The problem of deciding whether a \onlytsatt{} instance is satisfiable or not is the \onlytsatt{} problem.
\end{dfn}

\noindent An example for a \onlytsatt{} instance  is $\varphi_1(x_1, x_2, x_3) = (x_1 \vee x_2 \vee x_3) \wedge (\neg x_1  \vee \neg x_2 \vee x_3)$. In this case $\varphi_1$ is satisfiable, as for example $\varphi(1, 0, 0) = 1$. \\ Instances of the \onlytsatt{} problem, as well as many other NP-hard problems, are subject to the phase transition phenomenon \cite{cheeseman1991really}. Problem classes that are subject to the phase transition phenomenon are at first easy to solve until their parameterization reaches a critical point from which on they are suddenly very hard to solve or even become very rapidly unsolvable. For randomly created \onlytsatt{} instances with literals uniformly drawn from the variable pool of the \onlytsatt{} instance, the critical point of the phase transition is reached when the quotient of the number of clauses ($m$) and the number of variables ($n$) is approximately $\frac{m}{n} = 4.24$ \cite{gent1994sat}.


\subsection{Quadratic Unconstrained Binary Optimization (QUBO)}
\noindent One way of solving problems on quantum computers is to transform them into an instance of QUBO, as QUBO is the accepted input for quantum algorithms like quantum annealing \cite{mcgeoch2020theory} and the QAOA algorithm \cite{farhi2014quantum} on quantum gate model computers. 

\begin{dfn}[QUBO \cite{glover2018tutorial}]
Let $\mathcal{Q} \in \mathbb{R}^{n \times n}$ be a square matrix and let $x \in \mathbb{B}^n$ be an $n$-dimensional vector of Boolean variables. The QUBO problem is given as:
$$
 \text{minimize} \quad H_{\textit{QUBO}}(x) = x^T\mathcal{Q}x = 
\sum_{i}^n\mathcal{Q}_{ii}x_i + \sum_{i <j}^n\mathcal{Q}_{ij}x_ix_j
$$
\end{dfn}
\noindent We call $H_{\textit{QUBO}}(x)$ the (QUBO) energy of vector $x$. Matrix $Q$ will also be called \emph{QUBO matrix} or just \emph{QUBO}. QUBO is closely related to the Ising spin glass problem (Ising), which is defined as follows:
\begin{equation}
   \text{minimize} \quad H_{\textit{Ising}}(s) = \sum_{i}h_is_i + \sum_{i < j}J_{ij}s_is_j 
\end{equation}
\noindent Here, $h$ is an $n$-dimensional real-valued vector, $J$ is a real-valued $n \times n$-dimensional upper triangular matrix, and $s_i \in \{-1, 1\}$ are spin variables. \\

\noindent QUBO and Ising are isomorphic. That means a QUBO instance can equivalently be expressed as an Ising instance and vice versa. To transform an instance of QUBO into an instance of Ising, one can use transformation $x_i = (s_i + 1)/2$ \cite{tanahashi2019application}. Because of this isomorphism, we will use QUBO and Ising interchangeably in this work. QUBO and Ising are both NP-hard problems \cite{glover2018tutorial}.


\subsection{Maximum Weight Independent Set (MWIS)} \label{sec:mwis}
\noindent In this paper we will use a \onlytsatt{}-to-QUBO transformation that requires an understanding of the Maximum Weight Independent Set (MWIS) problem. Thus, we will review the necessary foundations here. For the remainder of this section, let $G = (V, E)$ be an undirected graph with vertex set $V$ and edge set $E$. \\
A subset $V' \subset V$ of the vertex set of $G$ is called an \emph{independent set}, if all vertices of $V'$ are pairwise non-adjacent, i.e., if $u, v$ are vertices in $V'$, it follows that $(u, v) \notin E$ \cite{lamm2019exactly}.

\begin{dfn}[Maximum Weight Independent Set]
In the MWIS problem, each vertex of $G$ gets assigned a weight. Let $w_v \in \mathbb{R^+}$ be the weight that is assigned to vertex $v$. The MWIS problem is to find an independent set $I$ of $G$, such that the total weight $w_{total} = \sum_{v \in I} w_v$ is the largest amongst all possible independent sets  \cite{lamm2019exactly}.
\end{dfn}
\noindent To solve the MWIS problems on quantum annealers, they need to be transformed to QUBO instances. This can be done as follows \cite{choi2010adiabatic}:

\noindent If $ Q_{ij} \ge |\min\{w_i,w_j\}| $ for all $(i, j) \in E$, then the maximum value of
$$
H(x_1, ..., x_n) =  -\sum_{i \in V} w_ix_i + \sum_{(i, j) \in E} Q_{ij}x_ix_j
$$
is the total value ($w_{total}$) of the MWIS. The variables $w_i$ denote the weight that was assigned to vertex $i$ and the variables  $x_i \in \{0, 1\}$ are binary variables. In particular, the maximum weight independent set of graph $G$ is given by $\{i \in V: x_i^* = 1\}$ where $(x_1^*, \dots, x_n^*) = \argmin_{(x_1, \dots, x_n)} H(x_1,\dots ,x_n)$.


\section{Related Work} \label{sec:soa_qubos}
\noindent In this section, we are going to review well-known \onlytsatt-to-QUBO transformations that will later be used as a baseline for our algorithmicly generated \onlytsatt-to-QUBO transformations as well as the concept of pattern QUBOs.


\subsection{Choi}
\noindent The first \onlytsatt-to-QUBO transformation we are going to review was published by Choi \cite{choi2010adiabatic}. In this approach, a \onlytsatt{} instance will be reduced to an instance of MWIS:\\\\
Given a \onlytsatt{} instance $\varphi(x_1, ..., x_n) = C_1 \wedge ... \wedge C_m$ with $n$ variables and $m$ clauses and an empty graph $G_{\textit{SAT}} = (V_{G_{\textit{SAT}}}, E_{G_{\textit{SAT}}})$.

\begin{itemize}
    \item For each clause $C_i = l_{i,1} \vee l_{i,2} \vee l_{i,3}$, we add a new 3-node clique to $G_{\textit{SAT}}$. The three nodes are labeled by the literals  $l_{i,1}, l_{i,2},$ and $l_{i,3}$ of clause $C_i$.
    \item A connection between nodes of different 3-node cliques will be added, if the labels of these nodes are in conflict. That is, iff $l_{i,s} = \neg l_{j,t}$ for clause indices $i \neq j$ and $s,t \in \{1,2,3\}$ we add edge $(l_{i,s}, l_{j,t})$ to $G_{\textit{SAT}}$.
\end{itemize}
In the resulting graph $G_{\textit{SAT}}$, every vertex gets weighted by the same weight $w_{\textit{vertex}} \in \mathbb{R}^+$, which can be chosen arbitrarily. This results in an MWIS instance in which the optimal solutions correspond to solutions of the input \onlytsatt{} instance. To solve this MWIS instance on a quantum annealer, it has to be transformed into an instance of QUBO as described in Sec. \ref{sec:mwis}.


\subsection{Chancellor}\label{sec:chancellor}
\noindent Chancellor \cite{chancellor2016direct} proposed different methods to transform \onlytsatt{} instances to Ising spin glass problems. In this paper, we will use the first proposed method in the ``special cases'' section of \cite{chancellor2016direct}. The underlying idea of this transformation is that for each clause of a \onlytsatt{} instance a new Ising spin glass problem will be constructed. All the created clause Ising spin glass problems will then be ``added'' together. We will now proceed to explain Chancellor's method as proposed in \cite{chancellor2016direct}:\\
Let $C_i = (x_1 \vee x_2 \vee x_3)$ be a clause of a \onlytsatt{} instance over the Boolean variables $x_1, x_2, x_3$.
If we set 1$:=$True and  0$:=$False, then the following holds:
\begin{equation}\label{eq:chancellor_1}
    (x_1 \vee x_2 \vee x_3) = x_1 + x_2 + x_3 -x_1x_2 -x_1x_3 - x_2x_3 + x_1x_2x_3
\end{equation}
This expression is maximized (the value is 1), whenever the clause is satisfied, and zero otherwise.
As we defined Ising (and QUBO) to be a minimization problem, we transform Eq. \ref{eq:chancellor_1} into the equivalent minimization version:
\begin{equation}\label{eq:chancellor_2}
   -(x_1 \vee x_2 \vee x_3) = -x_1 - x_2 - x_3  + x_1x_2  + x_1x_3 + x_2x_3 - x_1x_2x_3
\end{equation}
Now, whenever the clause is satisfied the value is $-1$ and zero otherwise. As Chancellor's approach uses spin variables, the Boolean variables $x_1, x_2, x_3$ in Eq. \ref{eq:chancellor_2} will be replaced by corresponding spin variables $s_1, s_2, s_3 \in \{-1, 1\}$ by using transformation $x_i = \frac{1}{2}(s_i + 1)$. This leads to:
\begin{equation}\label{eq:chancellor_3}
    \begin{aligned}
        -(x_1 \vee x_2 \vee x_3) =
        -\frac{1}{8}s_1 -\frac{1}{8}s_2 -\frac{1}{8}s_3  + \frac{1}{8}s_1s_2 +  \\ +\frac{1}{8}s_1s_3  + \frac{1}{8}s_2s_3 - \frac{1}{8}s_1s_2s_3- \frac{7}{8}
    \end{aligned}
\end{equation}
We will call the right hand side of Eq. \ref{eq:chancellor_3} the \emph{spin representation of a clause}.
The goal now is to create an Ising sping glass problem in which all the assignments to Boolean variables $x_i$ (or equivalently to spin variables $s_i$) that satisfy the clause have the exact same minimal energy, while the one assignment that does not satisfy the clause has a higher energy. We observe that the rightmost term  in Eq. \ref{eq:chancellor_3} that involves spin variables is cubic, but Ising and QUBO can only model quadratic interactions natively. Chancellor showed that this cubic term can nevertheless be expressed in Ising, by adding an additional spin $s_a$, also called \emph{ancilla} spin. The cubic term is modeled as follows:
\begin{equation} \label{eq:chancellor_clause}
    \begin{aligned}
I_{\textit{cubic}}(s_{1}, s_{2}, s_{3}) = J \sum_{i=1}^3\sum_{j=1}^{i-1}c(i)c(j)s_{i}s_{j} + h \sum_{i=1}^3c(i)s_{i} + \\ J_a \sum_{i=1}^3c(i)s_{i}s_{a} + h_{a}s_{a}
    \end{aligned}
\end{equation}
The terms $c(i)$ represent the sign of the corresponding Boolean variable $x_i$ in the clause we are trying to implement. As all the variables in the clause of Eq. \ref{eq:chancellor_3} are not negated, $c(i) = 1$ for all $i$. Parameters $J, J_a, h, h_a$ need to satisfy the following constraints:
\begin{itemize}
    \item $h$ is the value of the coefficient of the cubic term in the spin representation of the clause
    \item $h_a = 2h$
    \item $J_a = 2J > |h|$, which means the magnitude of $J$ can be chosen freely, as long as $2J > |h|$ holds.
\end{itemize}
We have now constructed an Ising spin glass representation of the cubic term. To get to the Ising spin glass representation of the clause we want to implement we add the coefficients of the linear and quadratic terms of the spin representation of the clause we are implementing (see Eq. \ref{eq:chancellor_3}) to the Ising representation of the cubic termin in Eq. \ref{eq:chancellor_clause}:
\begin{equation}
    \begin{aligned}
        I_{\textit{clause}} = I_{\textit{cubic}} + l_{1}s_1 + l_2s_2 + l_3s_3 + \\ q_{12}s_1s_2 + q_{13}s_1s_3 + q_{23}s_2s_3 
    \end{aligned}
\end{equation}
Terms $l_i, q_{ij} \in \{+1, -1\}$ are the coefficients of the corresponding linear and quadratic terms of Eq. \ref{eq:chancellor_3}.\\
This procedure is repeated for all clauses of the given \onlytsatt{} instance. Finally, we want to construct the Ising representation of the \onlytsatt{} instance by ``combining'' all of the Ising representations of the clauses. That is, we are going to superimpose the construction of the individual clauses on a common set of logical spin variables. To do so, suppose a given \onlytsatt{} formula consists of variables $x_1, ..., x_n$ and $m$ clauses $C_1, ... , C_m$. The corresponding Ising minimization problems will consist of spin variables $s_1, ..., s_n$, which directly correspond to Boolean variables $x_1, ..., x_n$ and $m$ additional ancilla qubits $s_{a_1}, ..., s_{a_m}$, which are needed to model the cubic term of each clause (see Eq. \ref{eq:chancellor_clause}).\\\\ Superimposing works as follows \cite{chancellor2016direct}:
\begin{itemize}
    \item For each $s_i$: Let $K$ be the set of all clauses in which $x_i$ or $-x_i$ appears as a variable. Then, $h_i = \sum_{l=1}^{|K|}h(i)_l$.
    The terms $h(i)_l$ are calculated by $h(i)_l = h \cdot c(i)_l + \textit{coeff}(s_i)_l$. Term $h \cdot c(i)_l$, where $c(i)_l$ is the sign of variable $x_i$ in the $l$-th clause of $K$, results from Eq. \ref{eq:chancellor_clause}. Term $\textit{coeff}(s_i)_l$ is the coefficient of the term of the spin variable $s_i$ in the spin representation of the $l$-th clause of $K$.
    \item For all pairs of spins $s_i$, $s_j$:  Let $P$ be the set of clauses in which variables $x_i$ and $x_j$ appear together, $i < j$: $J_{ij} = \sum_{l=1}^{|P|}J(ij)_l$. Terms $J(ij)_l$ are calculated by $J(ij)_l = J \cdot c(i)_lc(j)_l + \textit{coeff}(s_is_j)_l$. Term $\textit{coeff}(s_is_j)_l$ is the coefficient of the quadratic term $s_is_j$ in the spin representation of the $l$-th clause of $P$.
    \item All quadratic and linear values that include an ancilla qubit will not be modified at all. 
\end{itemize}

\noindent As Ising and QUBO are isomorphic, the resulting Ising representation of a given \onlytsatt{} instance can be transformed into an equivalent QUBO instance. As the remainder of this work will only use QUBO instances, we will refer to Chancellor's approach as a \onlytsatt{}-to-QUBO method, by implicitly using the isomorphism between Ising and QUBO.\\\\ As the mechanism of superimposing will be heavily used in the remainder of this paper, we illustrate Chancellor's approach and the method of superimposing in a simple example.

\begin{exmp}
Suppose we are given \onlytsatt{} instance $\varphi(x_1, x_2, x_3, x_4) = (x_1 \vee x_2 \vee x_3) \wedge (x_1 \vee \neg x_2 \vee x_4)$. For each of the clauses we construct an Ising problem that corresponds to the clause as explained previously. To do so, we use  $J = |h| = \frac{1}{8}$. We then transform the resulting clause Ising problems into an instance of QUBO. The resulting QUBOs are shown in Table \ref{example:chancellor}.
\end{exmp}

\begin{table}[!htb]
\centering
    \caption{QUBOs for \onlytsatt{} clauses  $(x_1 \vee x_2 \vee x_3)$ and $(x_1 \vee \neg x_2 \vee x_4)$ }
    \begin{subtable}{.5\linewidth}
      
         \caption{QUBO for clause \\ $(x_1 \vee x_2 \vee x_3$)}
        \begin{tabular}[h]{|r||p{0.3cm}|p{0.3cm}|p{0.3cm}|c|}
\hline
& $x_1$ & $x_2$ & $x_3$ & $A_1$ \\
\hline
\hline
$x_1$ & \textcolor{orange}{-2} & \textcolor{orange}{1} & \textcolor{orange}{1} & \textcolor{orange}{1} \\
\hline
$x_2$ &    &\textcolor{orange}{-2} & \textcolor{orange}{1} & \textcolor{orange}{1} \\
\hline
$x_3$ &    &   &\textcolor{orange}{-2} & \textcolor{orange}{1} \\
\hline
$A_1$ &    &   &   & \textcolor{orange}{-2} \\
\hline
\end{tabular}
  \label{example_chancellor_quboa}
    \end{subtable}%
    \begin{subtable}{.5\linewidth}
    
        \caption{QUBO for clause\\ $(x_1 \vee \neg x_2 \vee x_4)$}
        \begin{tabular}[h]{|r||p{0.35cm}|p{0.35cm}|p{0.35cm}|c|}
\hline
& $x_1$ & $x_2$ & $x_4$ & $A_2$ \\
\hline
\hline
$x_1$ & -1 & 0 & 1 & 1 \\
\hline
$x_2$ &    & 0 & 0 & 1 \\
\hline
$x_4$ &    &   &-1 & 1 \\
\hline
$A_1$ &    &   &   & -1 \\
\hline
\end{tabular}
\label{example_chancellor_qubob}
    \end{subtable} 
     \label{example:chancellor}
\end{table}

\noindent Variables $A_1$ and $A_2$ in Table \ref{example:chancellor} are ancilla variables for each of the clauses. Note that although variable $x_2$ is negated in clause $(x_1 \vee \neg x_2 \vee x_4)$, we use $x_2$ as the QUBO variable in the resulting clause QUBO (shown in Table \ref{example_chancellor_qubob}). By following the process of superimposing, one receives the QUBO shown in Table \ref{example:chancellor_2}. We colored the entries of the QUBO shown in Table \ref{example:chancellor_2}, so one can easily understand the process of superimposing visually. All orange colored values in Table \ref{example:chancellor_2} are taken from the clause QUBO for clause ($x_1 \vee x_2 \vee x_3)$ (shown in Table \ref{example_chancellor_quboa}) and all the black values are taken from the clause QUBO for clause $(x_1 \vee \neg x_2 \vee x_4)$ (shown in Table \ref{example_chancellor_qubob}).

\begin{table}[!htb]
\centering
    \caption{QUBOs for \onlytsatt{} clauses   $(x_1 \vee x_2 \vee x_3)$ and $(x_1 \vee \neg x_2 \vee x_4)$ }

        \begin{tabular}[h]{|r||p{0.7cm}|p{0.7cm}|p{0.7cm}|p{0.7cm}| p{0.3cm}|c|}
\hline
& $x_1$ & $x_2$ & $x_3$ & $x_4$& $A_1$ & $A_2$ \\
\hline
\hline
$x_1$ & \textcolor{orange}{-2} - 1& \textcolor{orange}{1} + 0 & \textcolor{orange}{1} & 1  &\textcolor{orange}{1} & 1\\
\hline
$x_2$ &    &\textcolor{orange}{-2} + 0 & \textcolor{orange}{1} & 0 & \textcolor{orange}{1} & 1 \\
\hline
$x_3$ &    &   &\textcolor{orange}{-2} &  & \textcolor{orange}{1} & \\
\hline
$x_4$ &    &   &   & -1  & & 1\\
\hline
$A_1$ &    &   &   &   &\textcolor{orange}{-2} &\\
\hline
$A_2$ &    &   &   &   & & -1\\
\hline
\end{tabular}
     \label{example:chancellor_2}
\end{table}


\section{Pattern QUBOs}\label{sec:pattern_qubos}
\noindent The core contribution of our paper is based on a concept that is implicitly used in Chancellor's \onlytsatt{}-to-QUBO formulation and more explicitly described in the construction of the $(n+m)$--sized \onlytsatt{}-to-QUBO transformation in \cite{nusslein2023solving}. In this section we will explicitly point this concept out, introduce the name \emph{pattern QUBO} for it and show its use, as our first contribution. In the next Section, we will propose an algorithmic method that uses pattern QUBOs to automatically create new \onlytsatt{}-to-QUBO transformations.\\

\subsection{QUBO Energy of a \onlytsatt{} assignment}
\noindent We start by examining the QUBO for clause $(x_1 \vee x_2 \vee x_3)$ (shown in Table \ref{example_chancellor_quboa}). We call this QUBO $Q_0$ in the remainder of this section. QUBO $Q_0$ can be used to find satisfying assignments for clause $(x_1 \vee x_2 \vee x_3)$. That is, each Boolean vector $\textbf{x} = (x_1, x_2, x_3, a_1)$ that minimizes $\textbf{x}^TQ_0\textbf{x}$ corresponds to a satisfying assignment of clause $(x_1 \vee x_2 \vee x_3)$. As $a_1$ is an ancilla variable that is not part of the set of variables of a given clause, for each assignment of Boolean values to variables $x_1, x_2, x_3$ there are two corresponding Boolean vectors in the QUBO minimization $\textbf{x}^TQ_0\textbf{x}$: $(x_1, x_2, x_3, a_1 = 0)$ and $(x_1, x_2, x_3, a_1 = 1)$. We now define the energy of an assignment of Boolean values to the variables of a \onlytsatt{} instance:

\begin{dfn}[Energy of an assignment]
Let $x_1, ..., x_n$ be the variables of a \onlytsatt{} instance and let Q be a QUBO instance that uses $m$ additional ancilla variables $a_1, ..., a_m$ to represent the given \onlytsatt{} instance as a QUBO instance. Given an assignment $x_1 = k_1, ..., x_n = k_n$, where $k_1, ..., k_n \in \{0, 1\}$, of Boolean values to variables $x_1, ..., x_n$, we define a column vector $\textbf{x}$ as $\textbf{x} = (k_1, ...,k_n, a_1, ..., a_m)$.  The  energy of an assignment of Boolean values to the variables $x_1, ..., x_n$ is given by:
\begin{equation}
    \mathcal{E}(x_1=k_1, ..., x_n=k_n) = \min \textbf{x}^TQ\textbf{x}
\end{equation}
Additionally, if QUBO $Q$ was used to calculate the energy  of an assignment $x_1=k_1, ..., x_n=k_n$, we will also say: ``The assignment $x_1=k_1, ..., x_n=k_n$ has energy $\mathcal{E}$ in $Q$''. With this, we emphasize that the specific QUBO $Q$ was used to calculate the energy for assignment $x_1=k_1, ..., x_n=k_n$.
\end{dfn}
\noindent The energy of an assignment is thus the value one obtains by choosing the values of the $m$ ancilla variables $a_1, ..., a_m \in \{0, 1\}^m$ such that the corresponding QUBO optimization process yields the lowest value.\\
As an example we calculate the energy for assignment $x_1 = x_2 = x_3 = 0$ by using clause QUBO $Q_0$. The energy of this assignment is given by $\mathcal{E}(0,0,0) = \min ( v_1^TQ_0v_1,v_2^TQ_0v_2) = \min (0, -1) = -1$. Hence, the assignment $x_1 = x_2 = x_3 = 0$ has energy $-1$ in $Q$. Here, $v_1$ is the vector defined as $v_1^T = (0,0,0, a_1 = 0)$ and $v_2$ is the vector defined as $v_2^T = (0, 0, 0, a_1 = 1)$. Next, we define a term that will heavily be used in the following sections.
\begin{dfn}[Clause QUBO]\label{def:clause_qubo}
Let $C$ be a clause of a \onlytsatt{} instance. Let $S_{\textit{SAT}}$ be the set of assignments of Boolean values to the variables of clause $C$ that satisfy clause $C$. A \emph{clause QUBO} is a QUBO for which each $s \in S_{\textit{SAT}}$ has the same minimal energy and for which the only assignment that does not satisfy clause $C$ has a higher energy than any of the assignments of $S_{\textit{SAT}}$.
\end{dfn}

\subsection{Pattern QUBOs}\label{subsec:patternqubo}
\noindent To introduce the concept of pattern QUBOs, we want to first point out the following (implicitly already used) observation. Suppose we are given the \onlytsatt{} instance of Example \ref{example:chancellor}, $\varphi(x_1, x_2, x_3, x_4) = (x_1 \vee x_2 \vee x_3) \wedge (x_1 \vee \neg x_2 \vee x_4)$. To calculate the energy of an assignment of Boolean values to variables $x_1, ..., x_4$, there are two equivalent ways:
\begin{itemize}
    \item[i)] Calculate the energy of $x_1 = 1, x_2 = 0, x_3 = 0$ for the first clause by using the QUBO shown in Table \ref{example_chancellor_quboa}. Calculate the energy of $x_1 = 1, x_2 = 0, x_4 = 0$ for the second clause by using the QUBO shown in Table \ref{example_chancellor_qubob}. Add both values.
    \item[ii)] Calculate the energy of $x_1 = 1, x_2 = 0, x_3 = 0, x_4 = 0$ by using the superimposed QUBO shown in Table \ref{example:chancellor_2}.
\end{itemize}
It can easily be verified that this is true for all possible assignments (i.e., by visually understanding the construction of the superimposed QUBO in Table \ref{example:chancellor_2}). Thus, we note that for each assignment the energy of the superimposed QUBO (shown in Table \ref{example:chancellor_2}) is equal to the sum of the energies of the respective assignments for the smaller QUBOs (shown in Table \ref{example_chancellor_quboa} and \ref{example_chancellor_qubob})  that were used to create the superimposed QUBO.\\\\ 
An important consequence of this observation is that any assignment that does not satisfy the \onlytsatt{} instance cannot possess the lowest energy value that is possible for the superimposed QUBO. That is the case because the clause QUBOs (shown in Table \ref{example_chancellor_quboa}, \ref{example_chancellor_qubob}) that were used to create the superimposed QUBO (shown in Table \ref{example:chancellor_2}) were constructed such that only satisfying assignments for that clause have the lowest energy. An assignment that does not satisfy the \onlytsatt{} instance does also not satisfy at least one clause. This assignment has a higher than optimal energy for this clause. As the energy of an assignment of the superimposed QUBO is the sum of the energies of the respective assignments for the smaller QUBOs that were used to created the superimposed QUBO, it follows that the energy for this non-satisfying assignment cannot be the lowest possible energy of the superimposed QUBO. Thus, as long as every clause is transformed into a corresponding clause QUBO (see Def. \ref{def:clause_qubo}), one can create a QUBO representation of a \onlytsatt{} instance, by superimposing over all the clause QUBOs.\\

\noindent For \onlytsatt{} problems specifically, we observe that each clause of a \onlytsatt{} instance has either exactly zero, one,  two, or three negated variables. By rearranging the clause, we can assume, without loss of generality, that the negated variables are always at the end of the clause. Thus, each \onlytsatt{} clause is represented by one of the following four types of clauses \cite{nusslein2023solving}:
\begin{itemize}
    \item Type 0 := $(a \vee b \vee c)$
    \item Type 1 := $(a \vee b \vee \neg c)$
    \item Type 2 := $(a \vee \neg b \vee \neg c)$
    \item Type 3 := $(\neg a \vee \neg b \vee \neg c)$
\end{itemize}

\noindent By using the previous argument, to create a new \onlytsatt{}-to-QUBO transformation, we only need to find one clause QUBO for each of the four clause types. We will use the term \emph{pattern QUBOs} for QUBOs that are clause QUBOs for one of the four types (type 0 -- 3) of clauses.
Now We would like to demonstrate why the term \emph{pattern QUBO} is justified:\\\\
A pattern QUBO for a type 0 clause is shown in Table \ref{table:type_0_clause_qubo_prototype}.
\begin{table}[!htb]
\centering
    \caption{Clause QUBO for type 0 clause   $(a \vee b \vee c)$ }

        \begin{tabular}[h]{|r||p{0.7cm}|p{0.7cm}|p{0.7cm}|p{0.7cm}| p{0.3cm}|c|}
\hline
& $a$ & $b$ & $c$ & $A$ \\
\hline
\hline
$a$ & -2 & 1 &1 & 1 \\
\hline
$b$ &  & -2 & 1 & 1\\
\hline
$c$&   &  & -2 & 1 \\
\hline
$A$ &    &   &   & -2 \\

\hline
\end{tabular}
     \label{table:type_0_clause_qubo_prototype}
\end{table}

\noindent If we want to transform a given \onlytsatt{} instance with $m$ clauses to a QUBO instance and we encounter a clause with zero negations, say $(x_1 \vee x_3 \vee x_5)$, we can substitute the variables $a, b, c$ in type 0 pattern QUBO (shown in Table \ref{table:type_0_clause_qubo_prototype}) by $x_1, x_3, x_5$. By repeating this process for all $m$ clauses (using the correct type 0 -- 3 pattern QUBO for the respective clauses) we create $m$ QUBO representations of individual  \onlytsatt{} clauses. As explained earlier, the resulting clause QUBOs can then be combined into a single QUBO that represents the whole \onlytsatt{} instance, by using the method of superimposing.\\

\noindent A certain \onlytsatt{}-to-QUBO transformation can thus be seen as an ordered tuple $(c_0, c_1, c_2, c_3)$, where $c_i$ is a pattern QUBO for clause type $i$ (for $0 \leq i \leq 3$). Note in particular that there exist multiple such tuples, which we will demonstrate in the coming section.\\


\section{Algorithmic Method}\label{sec:algorithm}
\noindent In this section we are going to present an algorithmic method that uses the concept of pattern QUBOs to create  new \onlytsatt{}-to-QUBO transformations automatically. We will also perform a case study, in which we solve \onlytsatt{} instances using both, state-of-the-art and created-by-algorithm \onlytsatt{}-to-QUBO transformations on D-Wave's quantum annealer Advantage\_system4.1.

\subsection{Algorithm Description}\label{sec:algorith}
\noindent As explained in Sec. \ref{subsec:patternqubo}, to create new \onlytsatt{}-to-QUBO transformations it suffices to find pattern QUBOs for type 0 -- 3 clauses. These pattern QUBOs can be reused and combined (by superimposing) to create a QUBO instance corresponding to a given \onlytsatt{} instance.

\noindent We will now present an algorithmic method that uses the blueprint for a pattern QUBO shown in Table \ref{table:type_0_blueprint} to create new pattern QUBOs for clause types 0 -- 3.

\begin{table}[!htb]
    \centering
    \caption{Blueprint for a pattern QUBO of any clause type}

        \begin{tabular}[h]{|r||p{0.7cm}|p{0.7cm}|p{0.7cm}|p{0.7cm}| p{0.3cm}|c|}
\hline
& $a$ & $b$ & $c$ & $A$ \\
\hline
\hline
$a$ & $Q_1$ & $Q_2$ & $Q_3$ & $Q_4$ \\
\hline
$b$ &  & $Q_5$ & $Q_6$ & $Q_7$\\
\hline
$c$&   &  & $Q_8$ & $Q_9$ \\
\hline
$A$ &    &   &   & $Q_{10}$ \\

\hline
\end{tabular}
     \label{table:type_0_blueprint}
\end{table}

\noindent This method has to find values for all $Q_i$, such that the resulting QUBO is a clause QUBO for the given clause type. As an input to this method we have to specify a set of values from which $Q_{i}$ can be chosen.  If this set is small enough, we can use exhaustive search to search all possible combinations of the 10 variables:

\begin{algorithmic}[1]
\Require Minimum value $min \in \mathbb{R}$, maximum value, $max  \in \mathbb{R}$, and step size $step \in \mathbb{R}$
\Require Upper triangular matrix Q. All values of Q ($Q_1$ to $Q_{10}$) should be initialized $min$.
\Procedure{Search clause QUBOs}{clause type}
\State FoundQUBOS = \{\}
\While{$Q_i <= max \;\; \forall Q_i$}
\If{IsClauseQUBO(Q)}
\State FoundQUBOS.insert(Q)
\EndIf
\State $Q_1 \leftarrow Q_1 + step$
\For{$i=1$ to $9$}
\If{$Q_i > max$}
\State $Q_i \leftarrow min$
\State $Q_{i+1} \leftarrow Q_{i+1} + step$
\Else
\State break
\EndIf
\EndFor
\EndWhile
\State \textbf{return} FoundQUBOS
\EndProcedure
\end{algorithmic}
Function \emph{IsClauseQUBO(Q)} in line 4 checks whether given QUBO $Q$ is a clause QUBO for the given clause type. That is, this function checks whether all satisfying assignments for that clause type have the same energy in Q, while the one assignment that does not satisfy the clause specified by the given clause type has a higher energy than any of the satisfying assignments.\\

\noindent By using this method and  $min = -1$, $max = 1$ and $step = 1$, we find 6 pattern QUBOs forclause type 0, 7 pattern QUBOs for clause type 1, 6 pattern QUBOs for clause type 2, and 8 pattern QUBOs for clause type 3. As any tuple $(c_0, c_1, c_2, c_3)$, where $c_i$ is a pattern QUBO for clause type $i$ ($0 \leq i \leq 3$), can be seen as a valid \onlytsatt{}-to-QUBO transformation, we have created a total of $2016 = 6 \cdot 7 \cdot 6 \cdot 8$ \onlytsatt{}-to-QUBO transformations. By decreasing $min$ and increasing $max$, even more combinations will be found.\\

\noindent Because we are using a new ancilla variable in each of the pattern QUBOs, all \onlytsatt{}-to-QUBO transformations created by this method result in QUBO matrices of dimensions $n+m$, where $n$ is the number of variables and $m$ is the number of clauses of a \onlytsatt{} instance. Thus, by choosing appropriate values for $min, max$, and $step$, this method can also produce previously published \onlytsatt{}-to-QUBO transformations of size $n+m$. Chancellor's \onlytsatt{}-to-QUBO transformation, for example, can be found with our method by using $min = -2$, $max = 1$, and $step=1$.

\subsection{Case Study using Quantum Hardware}\label{sec:case_study}

\noindent We are now going to conduct a case study in which we show that our method can create competitive \onlytsatt{}-to-QUBO transformations  by comparing the results of a \onlytsatt{}-to-QUBO transformation created by our algorithm with the results of well known state-of-the-art \onlytsatt{}-to-QUBO transformations when solving \onlytsatt{} instances on D-Wave's quantum annealer \emph{Advantage\_system4.1}. That is, we will show, that our automatically created \onlytsatt{}-to-QUBO transformations can solve equally many \onlytsatt{} instances as state-of-the-art \onlytsatt{}-to-QUBO transformations and that they can find correct solutions with an equal or higher probability than state-of-the-art \onlytsatt{}-to-QUBO transformations.\\

\noindent To do so, we create 1000 \onlytsatt{} instances according to the following description:
\begin{itemize}
    \item Each \onlytsatt{} instance has $m = 50$ clauses and $n = 12$ variables ($m/n \approx 4.16$). 
    \item For each of the $m$ clauses, we uniformly select three different variables from the $n$ variables of the instance. Additionally, there is a 50\% chance that each of the three chosen variables will get negated in that clause.
    \item Verify that the created \onlytsatt{} instance is solvable (using a SAT solver)
\end{itemize}

\noindent As explained in Sec. \ref{sec:algorith}, by using our method with parameters $min = -1$, $max = 1$, and $step = 1$, we can create 2016  \onlytsatt{}-to-QUBO transformations. As we cannot evaluate all of these transformations on real quantum hardware in a meaningful study (for cost reasons), we selected one at random. We call this transformation the \emph{Algorithm QUBO} in the following evaluation.

\noindent We now apply Chancellor's, Choi's and the Algorithm's \onlytsatt{}-to-QUBO transformation to all of the formerly created 1000 \onlytsatt{} instances. Each of the resulting 3000 QUBO instances will be solved 1000 times on D-Wave's quantum annealer \emph{Advantage\_system4.1}. The results of this experiment are shown in Table \ref{table:algorithm_results_1}.

\begin{table}[!h]
\caption{Results of the case study. 1000 \onlytsatt{} instances were solved 1000 times by Chancellor's, Choi's, and the Algorithm's \onlytsatt{}-to-QUBO transformation.}
\label{table_example}
\centering
\begin{tabular}{l | c | c }
\hline
\bfseries QUBO & \bfseries \#solved instances & \bfseries \#correct solutions \\
\hline\hline
Algorithm QUBO & 963 (96.3\%) & 181614 ($\approx$ 18.16\%)\\
Chancellor & 958 (95.8\%) & 139168 ($\approx$ 13.91\%)\\
Choi & 792 (79.2\%) & 45255 ($\approx$ 4.53\%) \\
\hline
\end{tabular}
\label{table:algorithm_results_1}
\end{table}
\noindent We can see that the Algorithm's \onlytsatt{}-to-QUBO transformation solves the most of the 1000 \onlytsatt{} instances, closely followed by Chancellor's \onlytsatt{}-to-QUBO transformation. However, the Algorithm's QUBOs finds approximately 30\% more correct solutions than Chancellor's QUBOs. We also observe, that the results of Choi's QUBOs are far behind the results of Chancellor's and the Algorithm's QUBOs. The latter observation is in accordance with the comparison of these two transformations in \cite{zielinski2023influence}, where it was also observed that Choi's QUBOs seem to perform significantly worse than Chancellor's QUBOs on similar \onlytsatt{} instances.\\
The results of this case study show that our algorithmic method can produce \onlytsatt{}-to-QUBO transformations that can compete with or even outperform state-of-the-art \onlytsatt{}-to-QUBO transformations with regards to the number of solved instances resp. the number of correct solutions. We note that in the benchmark study performed in \cite{zielinski2023influence} the worst performing \onlytsatt{}-to-QUBO transformation produced QUBO instances of size $n+m$. As our algorithmic method also creates $n+m$ sized \onlytsatt{}-to-QUBO transformations, the worst performing \onlytsatt{}-to-QUBO transformation of the benchmark study in \cite{zielinski2023influence} can also be found with our algorithmic method. Thus, it should be clear that not every \onlytsatt{}-to-QUBO transformation created by our proposed algorithmic method will outperform state-of-the-art \onlytsatt{}-to-QUBO transformations.\\
Although it is expected that not every algorithmically created \onlytsatt{}-to-QUBO transformation can compete with state-of-the-art \onlytsatt{}-to-QUBO transformations (in terms of number of the probability of finding a correct solution to a \onlytsatt{} instance), we opened up the possibility of exploring the space of possible \onlytsatt{}-to-QUBO transformations and showed, that our method can find \onlytsatt{}-to-QUBO transformations, that can compete with state-of-the-art \onlytsatt{}-to-QUBO transformations.

\section{Approximate \onlytsatt{}-to-QUBO transformations}\label{sec:approximate_3sat}
\noindent In this section we will demonstrate another use of our proposed algorithmic method. That is, we will introduce \emph{approximate \onlytsatt{}-to-QUBO transformations}. These transformations sacrifice optimality in some cases, but need significantly fewer logical variables and thus also fewer qubits on physical hardware in return. We will demonstrate that these transformations can be surprisingly effective (i.e. they can find better solutions to some \onlytsatt{} problems than non-approximate state-of-the-art \onlytsatt{}-to-QUBO transformations).\\

\noindent A \onlytsatt{} clause consisting of three different variables has $8 = 2^3$ possible assignments of Boolean values to the three variables of the clause. Of those 8 assignments, 7 satisfy the clause and one does not. Thus, 7 assignments should have the same lowest energy in a clause QUBO while the 8th assignment should have a higher energy. All previous \onlytsatt{}-to-QUBO transformations created by our proposed algorithm fulfill these constraints. However, instead of requiring all 7  satisfying assignments to have the same lowest energy in a QUBO, we will now only require 6 of the 7 satisfying assignments to have the same lowest energy in a QUBO. Thus, one satisfying and one non-satisfying assignment will have a higher than optimal energy in this QUBO. By using this relaxed requirement, we sacrifice some correctness. That is, some satisfying assignments can no longer be found be finding the minimal energy for a QUBO constructed like this. There may even be cases in which an assignment, that has the lowest energy of a QUBO, does no longer correspond to a correct \onlytsatt{} solution. Nevertheless, we will demonstrate that this may be a worthwhile trade-off in some cases. We will now define approximate \onlytsatt{}-to-QUBO transformations.\\

\begin{dfn}[Approximate clause QUBO]\label{def:clause_qubo}
Let $C$ be a clause of a \onlytsatt{} instance. Let $S_{\textit{SAT}}$ be the set of assignments of Boolean values to the variables of the clause $C$ that satisfy clause $C$. Note that $|S_{\textit{SAT}}| = 7$, as there are 7 satisfying assignments for a single clause. An \emph{approximate clause QUBO} is a QUBO in which any 6 of the 7 elements of $S_{SAT}$ have the same minimal energy. The single non-satisfying assignment for that clause, as well as the remaining satisfying assignment, need to have a higher than minimal energy.
\end{dfn}

\noindent We will call a  \onlytsatt{}-to-QUBO transformation that consists of one or more approximate clause QUBOs an \emph{approximate \onlytsatt{}-to-QUBO transformation}.

\noindent To create approximate \onlytsatt{}-to-QUBO transformations with our proposed algorithmic method, we will use the new blueprint for pattern QUBOs shown in Table \ref{table:type_0_blueprint_approx}.

\begin{table}[!htb]
    \centering
    \caption{Blueprint for approximated pattern QUBOs}

        \begin{tabular}[h]{|r||p{0.7cm}|p{0.7cm}|p{0.7cm}|p{0.7cm}| p{0.3cm}}
\hline
& $a$ & $b$ & $c$  \\
\hline
\hline
$a$ & $Q_{1}$ & $Q_{2}$ & $Q_{3}$  \\
\hline
$b$ &  & $Q_{4}$ & $Q_6$  \\
\hline
$c$&   &  & $Q_{6}$  \\

\hline
\end{tabular}
     \label{table:type_0_blueprint_approx}
\end{table}

\noindent In comparison to the previously used blueprint for pattern QUBOs (shown in Table \ref{table:type_0_blueprint}), this blueprint only contains the three variables of a \onlytsatt{} clause, without the added ancilla variable $A$. We will now use our proposed algorithmic method to find approximate pattern QUBOs for the previously defined clause types 0 -- 3. To do so, we modify line 4 in our algorithmic method. Instead of method \emph{IsClauseQUBO(Q)}, we will now use method \emph{IsApproximateClauseQUBO(Q)}. The latter method checks whether 6 of the 7 possible satisfying assignments have equally lowest energy in $Q$, while the 2 remaining assignments must have a higher energy.\\

\noindent By using this modified version of our algorithm (and $min = -1$, $max=1$, and $step =1$), we find exactly four approximate pattern QUBOs for each of the clause types 0 -- 3. This amounts to a total of $4 \cdot 4 \cdot 4 \cdot 4 = 256$ approximate \onlytsatt{}-to-QUBO transformations. We select one at random and call this transformation \emph{Approx 1}. The approximate pattern QUBOs for type 0 -- 3 clauses are shown in Table \ref{table:approx1_clause_qubos}.

\begin{table}[!htb]
\centering
    \caption{Approximate pattern QUBOs for type 0 -- 3 clauses of the \emph{Approx 1} \onlytsatt{}-to-QUBO transformation}
    \begin{subtable}{.3\linewidth}
      
         \caption{Type 0 \& 3}
         \begin{tabular}[h]{|r||c|c|c|}
\hline
& $a$ & $b$ & $c$  \\
\hline
\hline
$a$ & -1 & 1 & 1  \\
\hline
$b$ &  & -1 & 1  \\
\hline
$c$&   &  & 1  \\

\hline
\end{tabular}
    \end{subtable}%
    \begin{subtable}{.3\linewidth}
    
        \caption{Type 1}
         \begin{tabular}[h]{|r||c|c|c|}
\hline
& $a$ & $b$ & $c$  \\
\hline
\hline
$a$ & -1 & 1 & 0  \\
\hline
$b$ &  & -1 & 0  \\
\hline
$c$&   &  &  0  \\

\hline
\end{tabular}
    \end{subtable} 
       \begin{subtable}{.3\linewidth}
       \caption{Type 2}
         \begin{tabular}[h]{|r||c|c|c|c|}
\hline
& $a$ & $b$ & $c$  \\
\hline
\hline
$a$ & 0 & 0 & 0  \\
\hline
$b$ &  & 0 & 1  \\
\hline
$c$&   &  &  0  \\

\hline
\end{tabular}
    \end{subtable}%
     \label{table:approx1_clause_qubos}
\end{table}
\noindent All \onlytsatt{}-to-QUBO transformations created this way result in a QUBO instance of dimensions $n \times n$ for a \onlytsatt{} instance with $n$ variables and $m$ clauses. This is significantly smaller than $(n+m) \times (n+m)$, which is the dimension of Chancellor's \onlytsatt{}-to-QUBO transformation, or the size of the \onlytsatt{}-to-QUBO transformations we created earlier with our algorithmic method.

\noindent Before we evaluate this transformation on quantum hardware, we want to introduce a second version of approximate \onlytsatt{}-to-QUBO transformations. In the second version of approximate \onlytsatt{}-to-QUBO transformations, we use approximate pattern QUBOs for only three of the four clause types 0 -- 3. For the remaining clause type, we use a correct pattern QUBO (i.e., with an additional ancilla variable). As an example for this idea, we create the second approximate \onlytsatt{}-to-QUBO transformation, which we call \emph{Approx 2} as follows: For clause types 0, 1, and 3, we use the approximate pattern QUBOs displayed in Table \ref{table:approx1_clause_qubos}. However, we replace the approximate type 2 pattern QUBO (displayed in Table \ref{table:approx1_clause_qubos}) with a correct pattern QUBO for the clause type 2 (shown in Table \ref{table:pattern_qubo_approx2}).

\begin{table}[h]
\centering
\caption{Pattern QUBO for clause type 2 in the \emph{Approx 2} \onlytsatt{}-to-QUBO transformation}
  \begin{tabular}[h]{|r||p{0.3cm}|p{0.3cm}|p{0.3cm}|c|}
\hline
& $x_1$ & $x_2$ & $x_3$ & $A$ \\
\hline
\hline
$x_1$ & 1 & -1 &  0 & -1 \\
\hline
$x_2$ &  & 0 &  0 & 1 \\
\hline
$x_3$  &  &  &  1 & -1 \\
\hline
$A$ &    &   &   & 0 \\
\hline
\end{tabular}
\label{table:pattern_qubo_approx2}
\end{table}
\noindent The \emph{Approx 2} \onlytsatt{}-to-QUBO transformation thus creates correct clause QUBOs for all type 2 clauses of a \onlytsatt{} instance and only approximates type 0, 1, and 3 clauses. In our construction of \onlytsatt{} instances (see Sec. \ref{sec:case_study}), each chosen variable of a clause has a 50\% chance to be negated. Thus, there is a chance of 37.5\% ($\frac{3}{8}$) that a created clause contains two negated variables. Thus, the \emph{Approx 2} \onlytsatt{}-to-QUBO transformation uses a correct clause QUBO for $\approx 37.5\%$ of its clauses, while approximating the remaining $\approx 62.5\%$. We will now solve the 1000 \onlytsatt{} instances that we created in Sec. \ref{sec:case_study} again on D-Wave's Advantage\_system4.1, using the approximate \onlytsatt{}-to-QUBO transformations \emph{Approx 1} and \emph{Approx 2}. The results of this experiment are shown in Table \ref{table:algorithm_results_approx}.
\begin{table}[!h]
\caption{Results of the case study. 1000 \onlytsatt{} instances were solved 1000 times each by the approximated \onlytsatt{}-to-QUBO transformations \emph{Approx 1} and \emph{Approx 2} }
\label{table_example}
\centering
\begin{tabular}{l | c | c }
\hline
\bfseries QUBO & \bfseries \#solved instances & \bfseries  \#correct solutions \\
\hline\hline
Approx 1 & 737 (73.7\%) & 102318 ($\approx$ 10.23\%)\\
Approx 2 & 749 (74.9\%) & 160320 ($\approx$ 16.03\%)\\
\hline
\end{tabular}
\label{table:algorithm_results_approx}
\end{table}

\noindent When comparing these results with the results of the previous case study for the non-approximate \onlytsatt{}-to-QUBO transformations (shown in Table \ref{table:algorithm_results_1}), we observe that both approximate \onlytsatt{}-to-QUBO transformations \emph{Approx 1 \& 2} solved the least amount of the 1000 \onlytsatt{} instances. However, both approximate \onlytsatt{}-to-QUBO transformations yield more correct solutions than Choi's transformation. Approx 1 yielded 2.26 times the number of correct solutions of Choi's transformation and Approx 2 yielded 3.5 times the number of correct solutions of Choi's transformation. In terms of total number of correct solutions, Approx 2 even outperformed Chancellor's transformation (Approx 2 yielded $\approx 15\%$ more correct answers). We will now show that this is not just a phenomenon for small \onlytsatt{} instances by conducting a second experiment on larger \onlytsatt{} instances.\\

\begin{figure}
\centering
    \includegraphics[width=0.5\textwidth]{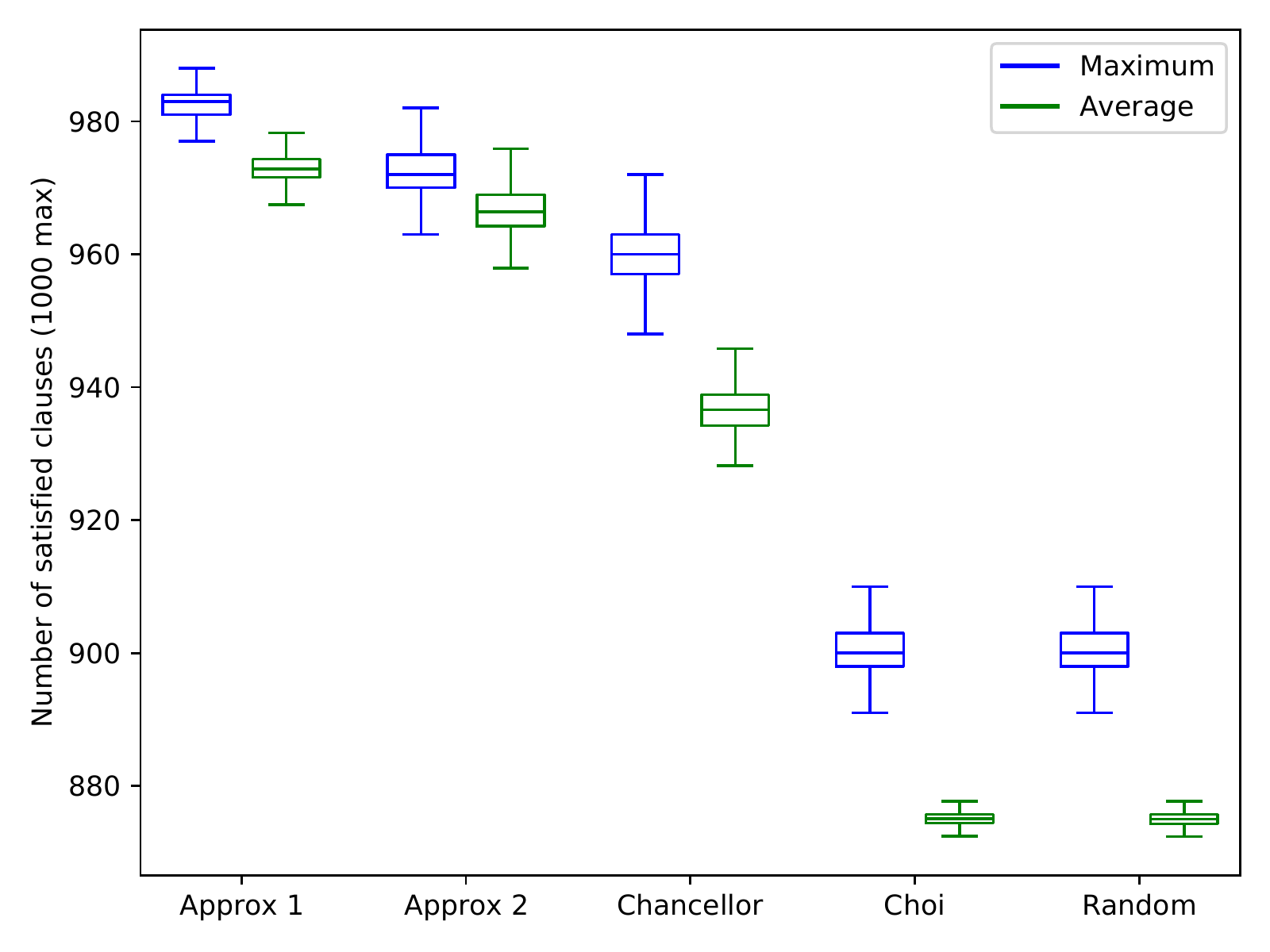}
    \caption{1000 \onlytsatt{} instances (240 variables, 1000 clauses) have been solved on classical hardware. Notably, the approximated \onlytsatt{}-to-QUBO transformations  (Approx 1 \& 2) outperform the established \onlytsatt{}-to-QUBO transformations Chancellor and Choi in both metrics.}
    \label{fig:approx_results_large}
\end{figure}

\noindent In this experiment, we create 1000 new \onlytsatt{} instances. Each of these new \onlytsatt{} instances consists of $n=240$ variables and $m=1000$ clauses. Each of these instances will be transformed into a QUBO instance using Chancellor's, Choi's, and the formerly created Approx 1 and Approx 2 \onlytsatt{}-to-QUBO transformations. All 4000 QUBOs will be solved 100 times each using D-Wave's Tabu Sampler. As in this experiment no instances were solved correctly by any of the QUBO transformations using the Tabu Sampler as an optimizer, we calculate the average number of \emph{clauses} solved and the maximum number of clauses solved for each \onlytsatt{}-to-QUBO transformation and for each formula and compare these values. The distributions of the average number of solved clauses and the maximum number of solved clauses per \onlytsatt{}-to-QUBO transformation and per formula are shown in  \figurename \ref{fig:approx_results_large}.

\noindent Surprisingly, both approximate \onlytsatt{}-to-QUBO transformations (Approx 1 \& 2) yield better results in both metrics than Chancellor's and Choi's \onlytsatt{}-to-QUBO transformation. As a reference we also included the results of randomly guessing assignments of Boolean variables 100 times for each of the 1000 \onlytsatt{} instances and calculating the same metrics for these assignments.\\
\noindent Another observation is that randomly guessing assignments seemed to be as effective as solving the \onlytsatt{} instances with the Tabu Sampler and the Choi \onlytsatt{}-to-QUBO transformation for these \onlytsatt{} instances. The probable cause for this is that the construction of Choi's transformation technically allows for contradictions to be introduced into answers of an optimizer (i.e., that a variable $x_i$ and its negation $\lnot x_i$ are simultaneously true). Although the occurrence of this is punished by an energy penalty in the construction of Choi's QUBOs, contradictions in the returned solutions are still not impossible. To the contrary, an analysis of Choi's results showed that all answers to every \onlytsatt{} instances contained over 180 contradictions (although there are only 240 variables). In the metrics mentioned above  we cannot count answers that contain such contradictions -- all clauses would be satisfied if we set each variable and its negation simultaneously to \emph{TRUE}. However, that is not a valid answer to a given \onlytsatt{} problem. To resolve these conflicts in Choi's transformation, we decided that if a variable $x_i$ and their negation $\neg x_i$ are set to \emph{True} in a returned answer, we set either $x_i$ or its negation to \emph{False} according to the following rule: If $x_i$ was assigned \emph{True} last, $x_i$ keeps its value and $\neg x_i$ will be set to \emph{False} (and vice versa). Thus, by fixing the assignment for approximately 75\% of the variables for every answer according to an arbitrary rule, we are essentially guessing. We conclude that Choi's QUBOs might not be suitable for such an analysis, but decided to keep it in the analysis, as we think that this realization is worth knowing about as well. \\
Finally, we want to highlight the significant reduction of needed variables (and thus physical qubits on quantum hardware). While the dimension of the QUBOs resulting from Chancellor's \onlytsatt{}-to-QUBO transformation is $n + m$, which in this case is 1240, the dimension of the QUBOs resulting from the Approx 1 transformation is $n$, which in this case is 240.\\
\noindent The results of this case study indicate that approximate \onlytsatt{}-to-QUBO transformations may be an interesting topic to study. Due to the approximative nature of these transformations it is to be expected, that there are types of \onlytsatt{} instances, for which approximate \onlytsatt{}-to-QUBO transformations are not feasible. Further studies need to be done, to identify these types of \onlytsatt{} instances.

\section{Conclusion}\label{sec:conclusion}
\noindent In this paper, we presented an algorithmic method that enables the automatic creation of new \onlytsatt{}-to-QUBO transformations. With this method, we increased the number of known \onlytsatt{}-to-QUBO transformations from approximately a dozen to many thousands. Our algorithmic method builds upon a concept that was already used implicitly in the construction of published \onlytsatt{}-to-QUBO transformations. We explicitly named this concept \emph{pattern QUBOs} and described how to use their properties in algorithmic methods to creating new \onlytsatt{}-to-QUBO transformations. Finally, we introduced the concept of approximate \onlytsatt{}-to-QUBO transformations. These transformations need significantly fewer variables (and thus significantly fewer qubits on quantum hardware) than known correct \onlytsatt{}-to-QUBO transformation, but also sacrifice optimality in some cases. Nevertheless, we demonstrated that approximate \onlytsatt{}-to-QUBO transformations can outperform currently known correct \onlytsatt{}-to-QUBO transformations in some cases.
\\Our work will enable the application of machine learning methods to create new \onlytsatt{}-to-QUBO transformations. Our proposed algorithmic method to finding new \onlytsatt{}-to-QUBO transformations is computationally expensive, as it is an exhaustive search. The use of evolutionary algorithms (like genetic algorithms) to find new pattern QUBOs in larger value ranges thus seems very promising. It is known that the chosen \onlytsatt{}-to-QUBO transformation can have an impact on the solution quality when solving a problem with quantum annealing. As we now have thousands of \onlytsatt{}-to-QUBO transformations, a machine learning framework that helps to find the best performing \onlytsatt{}-to-QUBO transformation amongst the thousands of available \onlytsatt{}-to-QUBO transformations for a given problem seems promising.
As approximate \onlytsatt{}-to-QUBO transformations turned out to be surprisingly effective in some cases, further studies should be performed to learn more about their properties.

\section*{Acknowledgment}
\addcontentsline{toc}{section}{Acknowledgment}
\noindent This paper was funded by the German Federal Ministry of Education and Research through the funding program ``quantum technologies --- from basic research to market'' (contract number: 13N16196).
\bibliographystyle{IEEEtran}
\bibliography{references}

\end{document}